\documentclass[aps,reprint,superscriptaddress,nofootinbib]{revtex4-1}
\usepackage{graphicx}
\usepackage{bm}
\usepackage[usenames,dvipsnames]{color}
\usepackage{amsmath}
\usepackage{amsfonts}
\usepackage{algpseudocode}
\usepackage{algorithm}
\usepackage{hyperref}

\renewcommand\vr[1]{\vec{r}_{#1}}

\begin{document}
\frenchspacing

\title{Non-metric interaction rules in models of active matter}

\author{Daniel M. Sussman}\email{daniel.m.sussman@emory.edu}
\affiliation{Department of Physics, Emory University, Atlanta, GA, USA}

\date{\today}

\begin{abstract}
This article is based on lecture notes for the \href{http://active-matter.eu}{Marie Curie Training school "Initial Training on Numerical Methods for Active Matter."} 
It is common in the study of a dizzying array of soft matter systems to perform agent-based simulations of particles interacting via conservative and often short-ranged forces. In this context, well-established algorithms for efficiently computing the set of pairs of interacting particles have established excellent open-source packages to efficiently simulate large systems over long time scales -- a crucial consideration given the separation in time- and length-scales often observed in soft matter. What happens, though, when we think more broadly about what it means to construct a neighbor list? What if interactions are non-reciprocal, or if the ``range'' of an interaction is determined not by a distance scale but according to some other consideration? As the field of soft and active matter increasingly considers the properties of living matter -- from the cellular to the super-organismal scale -- these questions become increasingly relevant, and encourage us to think about new physical and computational paradigms in the modeling of active matter. In this chapter we examine case studies in the use of non-metric interactions.
\end{abstract}

\maketitle

\section{Introduction}
In this chapter, we are going to think about new phenomena that emerge as soft matter moves from the study of squishy equilibrium materials  to thinking about active and sometimes living ``material'' systems. To connect with what we've seen so far in this series, we'll begin by thinking first \emph{not} about this physics, but about generalizations of some of the algorithms we use to efficiently perform simulations. We'll see that these algorithmic generalizations have natural analogs in physical systems. Here we go!

Agent-based numerical modeling of soft matter systems have been enormously influential in the development of the field of soft matter. A number of powerful, flexible, open-source packages for  performing large-scale molecular dynamics (MD)  simulations are readily available and continue to be developed \cite{plimpton1995fast,phillips2005scalable,van2005gromacs,anderson2008general,schoenholz2019jax}, helping radically lower the barrier to entry for researchers in the field. 

At its most basic level, the essential computational pattern in a molecular simulation is sketched in Alg. \ref{alg:MD}: A system of $N$ ``particles'' (colloids, soft spheres, interacting Lennard-Jones units, etc.) are initialized in a simulation domain equipped with a set of boundary conditions, and a sequence of time steps is executed. At each time step, given the current state of the system, all interparticle forces are computed and the information about these forces is used to forward integrate an equation of motion -- Newton's equations of motion, say, or the Langevin equation, and perhaps coupling the system degrees of freedom with a thermostat or barostat to perform simulations that appropriately sample different statistical ensembles \cite{frenkel2001understanding}. Already we are assuming that the forces we are interested in are simple, isotropic, pairwise, conservative ones -- this situation arises quite naturally in many physical systems, but there are certainly soft matter systems that require going beyond such assumptions!

\begin{algorithm}[H]
\caption{Fundamental MD loop}\label{alg:MD}
\begin{algorithmic}
\State Initialize system, $\vec{X}$
\For {Desired number of timesteps}
\State Evaluate all pairwise forces, $\vec{F}$
\State Update $\vec{X}$ based on $\vec{F}$
\EndFor
\end{algorithmic}
\end{algorithm}

One reason that these simulation techniques have been so powerful in the study of soft matter is that they are readily parallelizable, allowing large simulations of many orders of magnitude in time to be performed on modern hardware. This is crucial, since many soft matter systems are characterized by widely separated time and length scales! For instance, a micron-scale colloid is orders of magnitude larger than the atoms in the solvent it interacts with, and the time scales of solvent motion are radically faster than the time scales of colloidal motion. Even in single-component systems this issue arises: in a melt of high-molecular-weight polymers -- often represented by chains of repulsive spheres connected by stiff springs -- the time scale of overall polymer motion can be much longer than the time scale associated with the vibration of adjacent units composing the polymer \cite{rubinstein2003polymer}. Thus, to faithfully integrate the equations of motion, soft matter simulations often involve very many very short individual time steps. 

Earlier in this series we learned that in a naive implementation of the fundamental pattern of Alg. \ref{alg:MD}, the most computationally expensive piece is the calculation of the forces. In particular, even the evaluation of all pairwise distances between particles requires $\mathcal{O}(N^2)$ operations, and for many of the simulations we want to do $N^2$ is already too many. If the interparticle interactions are truly long-ranged, a variety of methods have been proposed that improve on the naive scaling, such as Ewald summation, particle-mesh methods, fast multipole methods \cite{frenkel2001understanding}. 

In many soft matter systems, however, it is common to care about potentials that have a finite range. In this setting, we can radically reduce the computational cost of building lists of neighbors. For instance, before computing distances between particles we can spatially partition the domain into a grid of ``cells'' and, during the force-computation step, only compute pairwise distances between points in nearby cells. If the simulation is of particles that all have roughly the same interaction range and are roughly  evenly  distributed throughout the domain, computing pairwise distances among particles that exert forces on each other becomes $\mathcal{O}(N)$. Not only is this version of the calculation computationally less expensive, it is also \emph{embarrassingly parallel} (a nice feature given the current state of computational hardware). Note, by the way, that for simulating systems with large disparities in particle sizes or densities other algorithms exist that similarly accelerate the simulation of short-ranged forces, such as using stenciled cell lists \cite{plimpton1995fast} or using a hierarchical tree-based structure to help partition the simulation domain rather than a grid of cells \cite{howard2019quantized}.

\section{Won't you be my neighbor?}
\begin{figure}[htbp]
\begin{center}
\includegraphics[width=0.9\linewidth]{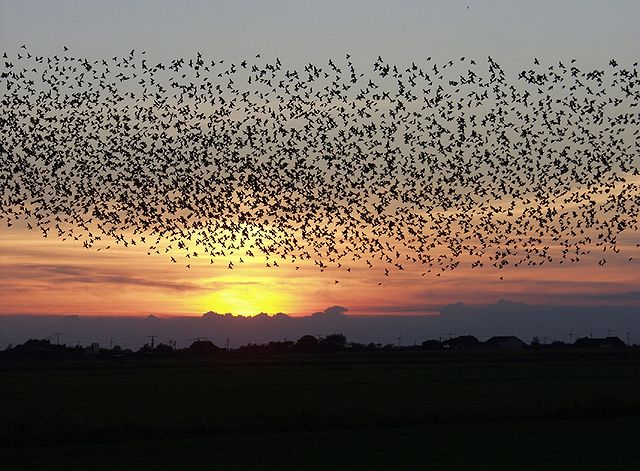}
\includegraphics[width=0.9\linewidth]{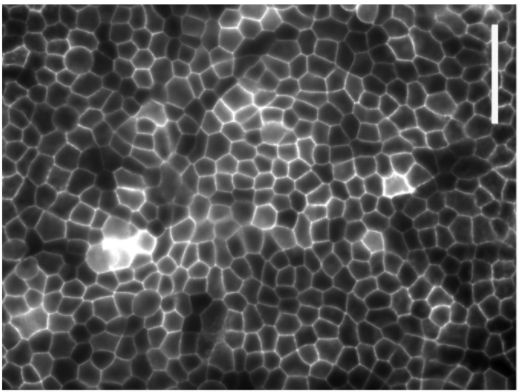}
\caption{
(Top) A  flock of starlings (Image from \href{http://www.pdfnet.dk}{http://www.pdfnet.dk} -- Material is in Public Domain).
(Bottom) Cross-sectional view of a monolayer of Madin-Darby Canine Kidney cells grown on a  collagen substrate \cite{devany2021cell}. Scale bar is 50 microns. }
\label{fig:neighbors}
\end{center}
\end{figure}

We spend so much of our time thinking about physical systems evolving according to conservative forces that this connection between \emph{neighbor lists} -- lists of interacting pairs of degrees of freedom -- and the interactions according to which the degrees of freedom exert forces is often assumed without much further thought. As active matter increasingly  looks to build highly coarse-grained models of living systems -- at the  level of cells or entire organisms, as shown in Fig. \ref{fig:neighbors} -- we should pause and question even these fundamental assumptions. 

For example, in the case studies below we will discuss classic models of flocking birds, in which each ``bird'' tries to align with its neighbors. Looking at a flock, or thinking about the neural processes involved in birds deciding which way to turn, it is far from clear that the ``neighbor list'' specifying which other birds a member of the flock is aligning with should be tied to strict distance cutoffs. Similarly, looking at a dense packing of cells, it seems clear that cell ``neighbors'' should be essentially determined by whether two cells share an edge -- an indirect  function of distance, perhaps, but one which has the character of a graph or network of interactions more than a distance-based cutoff. 

These are just two of a broader class of systems that we might say interact ``topologically,'' that  is, in which we define neighbors via something other than a strict  metric criteria. Generalizing our notion of neighbor relations takes us to the forefront of some very exciting current research in the soft active matter community, ranging from the new phenomena and scaling that can be found in such non-metric models \cite{ginelli2010relevance,sussman2018soft} to the role and importance of non-reciprocal forces \cite{you2020nonreciprocity,dadhichi2020nonmutual,ivlev2015statistical,lisin2020experimental}.

\section{Metric and topological flocking models}
The Vicsek model \cite{vicsek1995novel} is one of the canonical models in the study of the active matter. The original form of it studied out-of-equilibrium point  particles evolving according to a simple dynamical rule, and showed that beautiful collective organizational states emerged. Many variations of the Vicsek model have been proposed and study, but for our purposes we will focus on the simple ``vectorial'' version of the model \cite{ginelli2016physics}.

The essence of the model is to posit the overdamped dynamics of $N$ self-propelled particles. Each particle has a position, $\vec{r}_r$, and an orientation, $\hat{n}_i$, and at  each time step the positions of the particles are updated by moving at a constant speed, $v_0$, in the direction of the particle's orientation. Computationally, in a simple Eulerian integration of the equations of motion, at each time step we update the position of particle $i$ according to 
\begin{equation}
\Delta \vec{r}_i = (v_0\Delta t) \hat{n}_i(t).
\end{equation}
So far, this is just an uninteresting gas of non-interacting particles, each moving at some velocity. The dynamical rule added to make this model both deeply fascinating and technically non-trivial is to say that particles tend to align their direction of motion with nearby particles. Thus, at  each time step we also update  the  orientation of particle $i$ by
\begin{equation}\label{eq:align}
\hat{n}_i(t+\Delta t) = \mathcal{N} \left( \eta \vec{\xi}_i^{\left. t\right.}+ \frac{1}{m_i^t}\sum_{\langle ij\rangle} \hat{n}_j(t)\right).
\end{equation}
In the above equation, $\eta$ is a parameter that characterizes the strength of the noise in the model, $\vec{xi}_i^t$ is a random unit vector delta-correlated in both time and particle index, the sum is over the $m_i^t$ neighbors, $j$, that particle $i$ has at time $t$, and $\mathcal{N}$ is an operator that simply normalizes its argument. More simply: to determine a particles new direction of motion, we average the current direction of the particle's neighbors, add a bit of vectorial noise to this direction, and normalize the result so that $\hat{n}_i$ stays a unit vector.

How, though, should we choose the set of neighbors that each particle attempts to align with at each time step? Early versions of the model proposed a straightforward and quite natural metric criterion: there is a distance scale associated with particle alignment, and so a parameter was added to the model. In a given timestep, all pairs of particles within some distance, $r_0$, would count as  neighbors, and particles farther apart would not. This solution feels quite  natural, and it also has the virtue of mapping perfectly onto some of the efficient computational techniques we have learned for dealing with finite-ranged interactions. With this choice, there is a high-noise regime in which the particles are in a disordered state, a low-noise regime in which there  is a phase transition to a polar flocking state, and an intermediate regime in which a ``banded'' phase is observed, characterized by both overall polar order and periodic traveling density bands \cite{ginelli2016physics}.

While this kind of distance threshold for alignment feels natural, \emph{is it how birds actually align with each other}? A beautiful body of literature has attempted to infer simple, phenomenological interaction rules for flocking birds, shoaling fish, and other collectively moving groups of organisms \cite{cavagna2010scale, herbert2011inferring, gautrais2012deciphering}, and a consistent theme seems to be that these natural metric thresholds are likely poor descriptions of the ``social forces'' that represent organismal-scale interactions in moving-group dynamics. Some brief self-reflection of your own behavior in, say, moving crowds of different densities might supply the intuition that this should probably not be a surprise. 

For flocks of birds, the \href{https://www.isc.cnr.it/research/topics/physical-biology/biological-systems/starflag-a-project-on-collective-animal-behaviour/}{StarFlag project} reconstructed full three-dimensional  positions and trajectories of flocks of thousands of birds flying above a railway station in Rome. Statistical analyses of these trajectories revealed what your intuition perhaps suggested: a typical bird interacts not with all birds up to some distance, but rather with its six or seven \emph{closest neighbors} -- regardless of how far away those closest neighbors actually are \cite{cavagna2010scale,bialek2012statistical,attanasi2014information}.

We should be quite open, then, to interpreting the ``sum over neighbors'' in Eq. \ref{eq:align} without being restricted to the kinds of distance-based cutoffs we are so used to. What if, instead, we replaced the fixed-distance criteria with a rule that each particle interacted with its $k$ nearest  neighbors (with $k$  a parameter)? Or the $k$ nearest neighbors within some angular ``field of view''? What about a rule that constructed some triangulation of the set of positions -- say, a Delaunay triangulation -- with neighbor relations chosen according to the set of edges in the triangulation? In all of these cases, the interactions are still ``local,'' but they are local in a topological rather than metric sense.

In pioneering work, Ginelli and  Chat\'{e} showed that these changes \emph{matter}, with implications (for instance) about the nature of the transition between the different dynamical phases of the model \cite{ginelli2010relevance}. It is equally clear that the different topological criteria are \emph{not equivalent  to each other}. For instance, in a Delaunay triangulation the neighbor relation is still reciprocal -- particles $i$ and $j$ share an edge, say, so each is a neighbor of the other. In contrast, $k$-nearest-neighbors need not be reciprocal; in the presence of density inhomogeneities particle $i$ and be among the $k$ nearest neighbors of particle $j$ without $j$ being among the $k$ nearest neighbors of $i$. This opens up a universe of possibilities, with the effects of making these different choices for neighbor relations on the structural or dynamical phases of a model both interesting and poorly understood.

\section{Computational tools: Voronoi neighbor lists}
In our exploration of topological neighbor relations, we focus on Delaunay-based neighbor lists in this chapter. They break away from strictly metric considerations, but they still maintain reciprocity of particle interactions; in this sense they are a natural first generalization  to consider. So, what \emph{is} a Delaunay triangulation of a given set of points? One definition is  that for points in a two-dimensional plane, it is a triangulation that maximizes the minimum angle of all triangles in the triangulation (that is, the minimum angle in the Delaunay triangulation is at  least as large as the minimum angle found in all possible other triangulations of the point set). Equivalently, the Delaunay triangulation can be defined as the triangulation with the \emph{empty circumcircle} property: it is the triangulation such that no point in the point set is inside the circumcircle of any  triangle in the triangulation (the circumcircle is simply  the circle that passes through all of the vertices of a given polygon). Higher-dimensional versions of the Delaunay triangulation can be defined by the appropriate generalization of this to the empty circumsphere (or circum-hypersphere) property.

\begin{figure}[htbp]
\begin{center}
\includegraphics[width=0.4\linewidth]{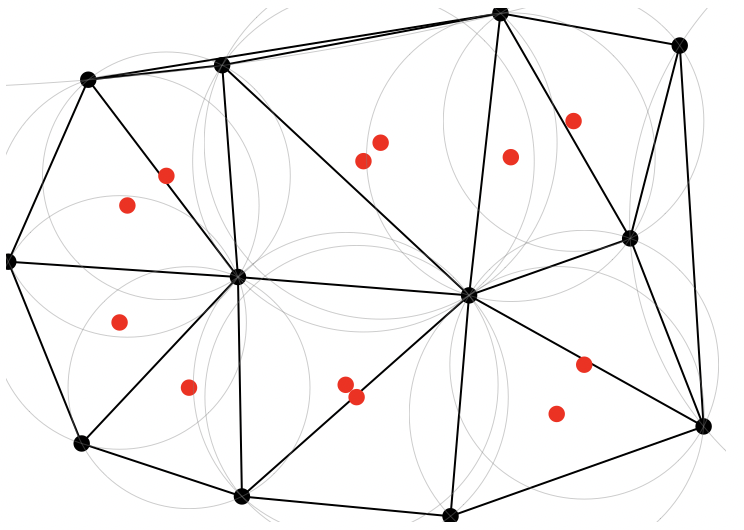} \includegraphics[width=0.4\linewidth]{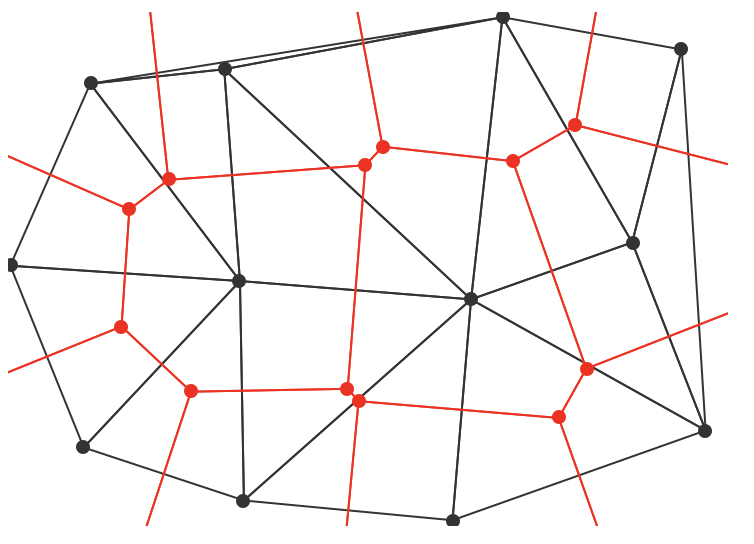}
\caption{
(Left) A Delaunay triangulation of a set of points (in black). Circumcircles of corresponding to each triangle are shown, with circumcircle centers in red. (Right) The Voronoi tessellation is dual to the Delaunay triangulation. Connecting the circumcircle centers of triangles that share an edge produces the Voronoi diagram of the original point set. }
\label{fig:delaunayVoro}
\end{center}
\end{figure}

An example Delaunay triangulation, along with the (empty) circumcircles of each triangle), is shown in Fig. \ref{fig:delaunayVoro}. The Delaunay triangulation is intimately related to the \emph{Voronoi diagram} of a point set, which is a partition of a space into polygonal (or polyhedral, or appropriate higher-dimensional generalizations) regions that are closer to a given point than to any other point in the set. In fact, the Voronoi diagram and the Delaunay triangulation are dual to one another: the vertices of the Voronoi diagram correspond to the circumcenters of the Delaunay triangulation, and two triangles share an edge in the Delaunay triangulation then the vertices corresponding to those circumcenters will be connected by an edge in the Voronoi diagram. This relationship is similarly  shown in  Fig. \ref{fig:delaunayVoro}.

The dual relation between Delaunay triangulations and Voronoi diagrams can often be exploited in computational settings. A wide variety of algorithms have been created to construct both types of graph, and if one has a Delaunay triangulation it is a trivial ($\mathcal{O}(N)$) operation to extract the Voronoi diagram (and vice versa). For instance, if one is simulating a Vicsek-style model where at each timestep every particle aligns with any particle in the first shell of its Voronoi neighbors \cite{ginelli2010relevance}, it might be computationally beneficial to avoid recomputing the entire Voronoi diagram if no change in the diagram occurred as a result of the most recent update of positions. It might be computationally convenient to formulate this condition -- ``do I need to recompute the Voronoi diagram?'' -- as a test  of the empty circumcircle property of the Delaunay triangulation.

These computational and algorithmic considerations can be quite important. Most MD algorithms exploit the easily parallelizable character of particles evolving according to short-ranged interactions to accelerate simulations (across multiple CPU cores, or using GPUs or TPUs), but most efficient algorithms for computing the edges and faces of Delaunay triangulations and Voronoi diagrams are fundamentally serial ones. In some ways, this is natural: even though, e.g., the Voronoi diagram can be thought of as a basically local characterization of the area around each point in a set, it must still obey global consistency conditions (e.g., in 2D the graph of the Voronoi diagram must obey Euler's formula for plane graphs).

This makes it more difficult to build efficient MD simulations in systems where neighbor relations are determined by such a criteria. One option is to have a subroutine in an otherwise highly parallelized simulation that makes a call to an efficient algorithm to construct the Voronoi diagram (or Delaunay triangulation) of the entire set  of particle positions in the simulation. Although scripting languages are not typically well-suited to large-scale simulations, existing packages can provide this functionality in Python, Matlab, Mathematica, and similar environments. On the high-performance side, a wrapper to quality open-source libraries that have been optimized for geometric calculations --  like the Computational Geometry Algorithms Library (CGAL) \cite{cgal:eb-19b,cgal:hs-ch3-19b} and Voro$++$ -- can be incorporated into your MD simulation software and called when needed. Recently, some work has gone into developing MD software packages that  incorporate these sorts of tessellations at a fundamental level, with a high level of parallelism exploited even in the triangulation protocol \cite{sussman2017cellGPU}. To date, most of these efforts have centered around efficiently simulating specific models of dense tissue, which we discuss next.

\section{Geometric models of dense tissue}
A natural setting in which this sort of tessellation-based neighbor list arises is in modeling dense, confluent tissue -- tissues in which there are no gaps between the cells composing the tissue. There is a rich history of looking at images of the sort seen in Fig. \ref{fig:neighbors} and taking the idea of ``cells as shapes'' seriously, representing the confluent tissue by polygonal or polyhedral tilings of space \cite{Honda1978, Honda2001, Hufnagel2006, Farhadifar2007, yan2019multicellular, alt2017vertex, noll2017active, popovic2020inferring}. These fundamentally geometrical models of cells have fascinating properties -- in some cases they display the sort of behavior one might expect given roughly \emph{any} reasonably-biologically-informed agent-based model in which each cell corresponds to just one or a handful of degrees of freedom, but in other cases they can be shown to support exotic mechanical states \cite{noll2017active,moshe2018geometric,kupferman2020continuum} and display unusual structural or dynamical scaling \cite{bi2016motility,teomy2018confluent}. To investigate these unusual properties, in recent years there have been substantial advances in the efficient numerical simulations of these and related agent-based models of dense tissue \cite{van2015simulating, sussman2017cellGPU, SAMOS, Fletcher2014, boromand2018jamming}.

To take a specific example, we consider using a computational package to generate the Voronoi diagram for a set of $N$ points in the plane. In the context of the sort of Voronoi-based neighbor list discussed above, it is natural to assign each Voronoi ``cell'' to its generating point: the location of the points will be the degrees of freedom of the model, and each point will have a particular geometric region associated with it at every point in time. Here we are thinking of a ``Voronoi model'' in the context of dense tissue, but related models have been used to also study foams, geological formation processes, and the partitioning of animal territorial domains \cite{weaire1984soap}. We proceed, as promised, to take the tessellation seriously and consider an energy functional that depends explicitly on the geometry of the tessellation:
\begin{equation}\label{eq:vmEnergy}
E = \sum_i^N\left( K_A (A_i-A_0)^2 + K_P(P_i-P_0)^2\right).
\end{equation}
Here $A_i$ and $P_i$  are the area and perimeter of cell $i$, $A_0$ and $P_0$ are ``preferred'' values of the area and perimeter for cell $i$, and $K_A$ and $K_P$ are area and perimeter moduli. In the context of dense tissue, the quadratic dependence on cell area can be interpreted as a cell monolayer's resistance to height fluctuations due to adhesions between cells and cell incompressibility, and the quadratic dependence on cell perimeter as a competition between active contractility of the actomyosin sub-cellular cortex and tension due to both cell-cell adhesions and cortical tension \cite{bi2016motility}. More generally, one might view this as subset of the possible low-order Taylor expansions that can be written in the various geometric properties of the tessellation. Indeed, in the absence of any area terms precisely this interpretation has been given, where the terms non-linear in perimeter are what separate out foam-like energy functionals from tissue-like energy functionals \cite{kim2018universal}. 

Since we will be simulating this model on a computer, it is convenient to non-dimensionalize the energy by choosing the unit of length to be $\sqrt{\langle A \rangle}$, where $\langle A \rangle$ is the average area of all of the cells in our simulation, letting $k_r = K_A \langle A \rangle/K_P$, and writing
\begin{equation}
\frac{E}{K_P \langle A \rangle} \equiv e = \sum_i k_r (a_i-1)^2 + (p_i-p_0)^2,
\end{equation}
where $a$ and $p$ refer to dimensionless areas and perimeters. This non-dimensionalization differs from Eq. \ref{eq:vmEnergy} only by an overall constant (assuming a finite domain that is completely occupied by  cells), and it makes clear that the density-dependence of the model enters via $p_0 = P_0/\sqrt{\langle A \rangle}$ -- an important point given the early interest of this model as one in which the jamming transition was one controlled by preferred shape rather than strict density considerations \cite{Bi2015,teomy2018confluent,merkel2018geometrically}

Now that we have a set of degrees of freedom (again, here, the location of points in 2D space), a rule for finding neighbors, and an energy functional (which depends on that neighbor-finding rule), we have all of the ingredients necessary to perform agent-based simulations. One could perform energy minimizations to find tissue ground states, use classic molecular dynamics algorithms to simulate them in $NVE$ or $NVT$ ensemble, or choose from any of a number of equations of motion corresponding to non-equilibrium, active-matter scenarios (self-propelled particle dynamics, or self-propulsion with noise, or chemotaxis, or...) -- the world is our oyster. All that is needed is an efficient way of taking the appropriate gradients of this energy functional in order to compute sets of forces acting on our degrees of freedom. For completeness, we show two variations of this calculation.

\subsection{Vertex model forces}
First, suppose we had a simpler version of our geometric model. Rather than taking a tessellation of a point set, suppose our degrees of freedom were actually the \emph{vertices} of a set of polygons tiling the plane. This sort of model -- often, unsurprisingly, referred to as a \emph{vertex model} -- has a similar flavor to Voronoi-based models, in that once again the neighbor list of our degrees of freedom comes \emph{not} from strictly distance-based considerations, but rather stems from a graph structure (i.e., from an explicit enumeration of vertex-vertex connections). Given such an enumeration, computing the forces on the vertices given the energy functional in Eq. \ref{eq:vmEnergy} is straightforward.

\begin{figure}[htbp]
\begin{center}
\includegraphics[width=0.45\linewidth]{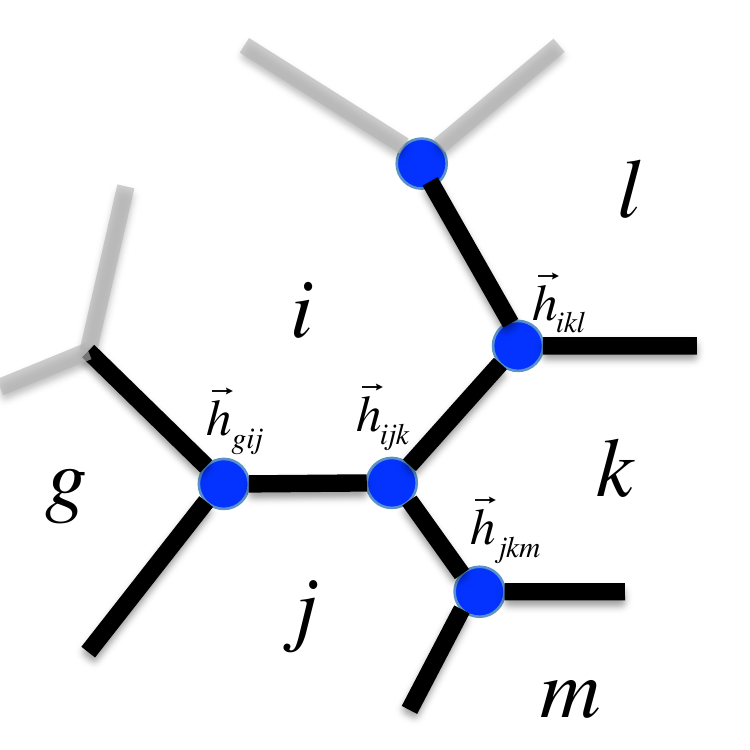}
\caption{Schematic diagram of cell $i$ and some of its neighbors, along with the associated Voronoi vertices (the circumcenters of three adjacent cells) labeled for convenient reference.}
\label{fig:forceSchematic}
\end{center}
\end{figure}

For reference, Fig. \ref{fig:forceSchematic} provides a schematic picture of a relevant patch of a two-dimensional tissue model, with each vertex labeled by the three cells it is adjacent to. From here, computing the forces is simply an exercise in patience / appropriate propagation of the chain rule. From the figure, since the motion of vertex $\vec{h}_{ijk}$ only changes the shape of cells $i,$ $j,$ and $k$, the force on vertex $\vec{h}_{ijk}$ is
\begin{equation}
-\frac{\partial E}{\partial \vec{h}_{ijk} } =-\left( \frac{\partial E_i}{\partial \vec{h}_{ijk} } + \frac{\partial E_j}{\partial \vec{h}_{ijk} } + \frac{\partial E_k}{\partial \vec{h}_{ijk} } \right).
\end{equation}
Each of the cell-specific energy derivatives are straightforward, e.g., 
\begin{equation}
\frac{\partial E_i}{\partial  \vec{h}_{ijk}} = 2 K_A(A_i-A_0)\frac{\partial A_i }{\partial  \vec{h}_{ijk}}+2 K_P(P_i-P_0)\frac{\partial P_i }{\partial  \vec{h}_{ijk}},
\end{equation}
where the area- and perimeter-derivatives with respect to vertex positions can themselves be written as follows. Let $\vec{t}_{ij} = \vec{h}_{ijk} -  \vec{h}_{gij}$, $\vec{t}_{ik} =   \vec{h}_{ikl}- \vec{h}_{ijk} $, with $\hat{t}_{ij}$ and $\hat{t}_{ik}$ being the unit vectors in those directions. Similarly, let $l_{ij}$ be the length of the edge between cell $i$ and $j$, and $\hat{n}_{ij}$ be the unit vector pointing outwardly normal to that cell edge. Then one can write
\begin{eqnarray}
\frac{\partial A_i }{\partial \vec{h}_{ijk}} &= &\frac{1}{2}\left( l_{ij} \hat{n}_{ij} +l_{ik} \hat{n}_{ik}   \right), \\
\frac{\partial P_i }{\partial \vec{h}_{ijk}} &= &-\left(\hat{t}_{ij}+\hat{t}_{ik}  \right).
\end{eqnarray}

\subsection{Voronoi model forces}
Returning to the class of models where the degrees of freedom are the positions of a point set we have tessellated, the same approach to computing forces can be used. Of course, the positions of the vertices of the Voronoi tessellation can be explicitly computed from the original point set, and so ultimately we are simply forced to carry out an extra layer of chain-rule calculations. Let's see this explicitly.

Following the notation in Bi et al. \cite{bi2016motility}, the force on cell $i$ in cartesian direction $\lambda$ can be computed as 
\begin{equation}
F_{i\lambda}=-\frac{\partial E}{\partial r_{i\lambda}} = -\frac{\partial E_i}{\partial r_{i\lambda}}-\sum_{<ij>}\frac{\partial E_j}{\partial r_{i\lambda}},
\end{equation}
the motion of cell $i$ changes the geometry, and hence the energy, of cell $i$ itself and all of its $\langle ij \rangle$ neighbors, $j$. Again using the configuration in Fig. \ref{fig:forceSchematic}, all of these terms can be expanded via the chain rule, for instance:
\begin{equation}
\frac{\partial E_k}{\partial r_{i\lambda}} = \sum_{\nu}\left(\frac{\partial E_k}{\partial h_{ijk,\nu}}\frac{\partial h_{ijk,\nu}}{\partial r_{i\lambda}} +\frac{\partial E_k}{\partial h_{ikl,\nu}}\frac{\partial h_{ikl,\nu}}{\partial r_{i\lambda}} \right).
\end{equation}
Here these are the only terms needed, since the other voronoi vertices associated with cell $k$ (the middle of the three neighboring cells in clockwise order) do not depend on the position of cell $i$. The partial derivatives depend on the positions of $\vec{h}_{jkm}$ and $\vec{h}_{kln}$, where $m$ is the cell other than $i$ that has both $j$ and $k$ as neighbors, and $n$ is the cell other than $i$ that has both $k$ and $l$ as neighbors (for this energy functional there is a dependence on the nearest and some of the next-nearest neighbors of cell $i$).

The derivative of the energy with respect to the vertices was calculated in the previous subsection, so the only additional piece that is needed are the derivatives of the Voronoi tessellation's vertices with respect to the position of the cell, e.g., $(\partial \vec{h}_{ijk})/(\partial \vr{i})$. These derivatives can be calculated efficiently as follows. We let $\vr{ij}$ denote the vector from $i$ to $j$, and define the auxiliary quantities
\begin{eqnarray}
c & = & \vec{r}_{ij,x}\vec{r}_{kj,y} - \vec{r}_{ij,y}\vec{r}_{kj,x}\\
d & = & 2 c^2\\
\vec{z} & = & \beta d \vr{ij} + \gamma d \vr{ik}\\
\beta d & = & -\left|\vr{ik}\right|^2 \cdot \left(\vr{ij}\cdot\vr{jk}\right)\\
\gamma d & = & \left|\vr{ij}\right|^2 \cdot \left(\vr{ik}\cdot\vr{jk}\right).
\end{eqnarray}

with $I_2$ representing the $2\times 2$ identity matrix and $\otimes$ the dyadic product, the desired change in Voronoi vertex position with respect to cell position is 
\begin{widetext}
\begin{equation}
\frac{\partial \vec{h}_{ijk}}{\partial \vr{i}} =I_2+ \frac{1}{d}\left[ \vr{ij}\otimes \left(\frac{\partial (\beta d)}{\partial \vr{i}}\right) + \vr{ik}\otimes \left(\frac{\partial (\gamma d)}{\partial \vr{i}}\right) - (\beta d+\gamma d)I_2 - \vec{z}\otimes \left( \frac{1}{d}\frac{\partial d}{\partial \vr{i}}\right)   \right],
\end{equation}
\end{widetext}
where
\begin{equation}
\frac{\partial (\beta d)}{\partial \vr{i}}  =  2 \left(\vr{ij}\cdot\vr{jk} \right)\vr{ik} +\left|\vr{ik}\right|^2 \vr{jk},
\end{equation}
\begin{equation}
\frac{\partial (\gamma d)}{\partial \vr{i}} = -2 \left(\vr{ik}\cdot\vr{jk} \right)\vr{ij} +\left|\vr{ij}\right|^2 \vr{jk}, 
\end{equation}
and
\begin{equation}
\frac{1}{d}\frac{\partial d}{\partial \vr{i}} =  \frac{2}{c} \left(-\vec{r}_{jk,y} \hat{x}+\vec{r}_{jk,x}\hat{y}  \right).
\end{equation}

\section{Discussion}

The point of the explicit calculation is not simply to recapitulate expressions that have appeared previously in the literature for the efficient calculation of vertex- and Voronoi-model forces \cite{bi2016motility,sussman2017cellGPU}, but to again emphasize the common character of these models with other agent-based simulations in the study of equilibrium and active matter. In this instance, the energy of a tissue configuration is dependent on the shape of the tessellated regions, and so in our computational work we need additional subroutines and data structures to calculate and keep track of these unusual sets of neighbor relations. Everything else, though, fits into the usual paradigm of a molecular dynamics simulation.

In the context of dense models of tissues, this perspective of starting with an unusual neighbor list and taking its implication seriously suggests a class of models with unusual behavior -- in  the way they can support stresses \cite{noll2017active}, in their ``low-temperature'' properties \cite{Sussman2018epl,sussman2018no}, in the way they can organize into collectively moving states  \cite{giavazzi2018flocking}, in the way they support ``curvotaxis'' \cite{PhysRevResearch.2.023417}, and so on. In the context of active models with less complex energy functionals, it is known that this ``topological'' way of choosing your neighbors changes the nature of the flocking transition  \cite{ginelli2010relevance}. Beyond this, though: an expanse of open questions!


\bibliography{NumericalMethods_DMS}

\end{document}